# The Role of Advanced Computer Architectures in Accelerating Artificial Intelligence Workloads

Shahid Amin [0009-0008-1130-0830] and Syed Pervez Hussnain Shah [0009-0005-3965-9093]

Department of Computer Science, Lahore Leads University

```
Email: Mr.shahidamin@gmail.com
```

```
Email: pervezhussnain@gmail.com
```

**Abstract.** The remarkable progress in Artificial Intelligence (AI) is foundationally linked to a concurrent revolution in computer architecture. As AI models, particularly Deep Neural Networks (DNNs), have grown in complexity, their massive computational demands have pushed traditional architectures to their limits. This paper provides a structured review of this co-evolution, analyzing the architectural landscape designed to accelerate modern AI workloads. We explore the dominant architectural paradigms Graphics Processing Units (GPUs), Application-Specific Integrated Circuits (ASICs), and Field-Programmable Gate Arrays (FPGAs) by breaking down their design philosophies, key features, and performance trade-offs. The core principles essential for performance and energy efficiency, including dataflow optimization, advanced memory hierarchies, sparsity, and quantization, are analyzed. Furthermore, this paper looks ahead to emerging technologies such as Processing-in-Memory (PIM) and neuromorphic computing, which may redefine future computation. By synthesizing architectural principles with quantitative performance data from industry-standard benchmarks, this survey presents a comprehensive picture of the AI accelerator landscape. We conclude that AI and computer architecture are in a symbiotic relationship, where hardware-software co-design is no longer an optimization but a necessity for future progress in computing.

Shahid Amin [0009-0008-1130-0830] and Syed Pervez Hussnain Shah

# 1      Introduction

Artificial Intelligence has entered an era of remarkable growth, touching everything from self-driving cars and medical diagnosis to creative arts and scientific research [1]. This progress, however, is built on a foundation of tremendous computational power [2]. The very algorithms driving these breakthroughs have a huge and rapidly growing appetite for processing power, memory, and data, creating a computational challenge that is fundamentally changing computer architecture [2].

The complexity of today's top AI models is growing much faster than Moore's Law ever predicted. Deep Neural Networks (DNNs), the heart of modern AI, have grown from models with thousands of parameters to massive foundation models, like Large Language Models (LLMs), with billions or even trillions of parameters [3]. Training a single one of these large models is a massive undertaking. It often requires thousands of Graphics Processing Units (GPUs) running for weeks or months, processing petabytes of data, and costing millions of dollars [3]. The challenge doesn't stop at training; deploying these models to serve millions of users with fast response times presents its own set of difficult computational problems[3]. This explosive growth in demand has made traditional computers inadequate, sparking a wave of architectural innovation.

For decades, the Von Neumann architecture which keeps the processor and memory separate has been the standard for computing [3]. While it worked well for general tasks, its main weakness, the "Von Neumann bottleneck," has become a major roadblock for data-heavy AI workload [4]. AI algorithms, especially DNNs, are constantly moving huge amounts of data (like model weights and activations) between memory and the processor. This constant back-and-forth takes up a lot of time and energy, often more than the actual computation itself [4].

General-purpose CPUs, which are designed for handling tasks one after another, are simply not a good fit for the kind of math AI requires [3]. The core operations in deep learning, like matrix multiplication, involve doing the same simple calculation on massive arrays of data at the same time. A CPU, with only a few powerful cores, can't handle this level of parallelism efficiently, leading to slow performance for any serious AI model [5].

To solve these problems, the industry has shifted toward domain-specific architectures (DSAs) hardware designed specifically for one job, in this case, AI [6]. These specialized chips, known as AI accelerators, are built from the ground up to match the unique computational patterns of neural networks [3].

This marks a major change from the one-size-fits-all approach of the past. AI accelerators use parallel processing, breaking large tasks into smaller pieces that can be run simultaneously across thousands of simpler cores [3]. This can be done in two main ways. The first uses separate hardware accelerators, like GPUs or custom ASICs, that work alongside a main CPU to handle the heavy lifting [6]. The second involves integrating AI acceleration capabilities directly into modern CPUs to provide a boost for specific functions, offering a more cost-effective solution for lighter tasks [6]. The rise of these accelerators, from tiny edge devices to huge cloud data centers, represents a paradigm shift in computer architecture, driven entirely by the needs of AI [2].

The evolution of computer architecture is no longer separate from the software it runs. Instead, advanced hardware and AI algorithms are now in a symbiotic co-evolution, where each one drives the other forward. This paper argues that understanding this relationship is key to understanding where AI and computing are today and where they are headed.

The primary contribution of this work is a holistic synthesis of the AI accelerator landscape, designed to bridge the gap between foundational academic theory and current industrial practice. While other excellent surveys exist, such as those by Del Core et al. on HPC accelerators and Zhang et al. on LLM hardware, our unique focus is on connecting core design principles such as dataflow, memory hierarchies, and quantization directly to the quantitative performance seen in the latest industry benchmarks. To achieve this, we provide a structured literature review, a comparative analysis of dominant architectural paradigms, and a discussion of open research challenges, offering a comprehensive guide for researchers and engineers in the field.

## 2      Literature Review

The field of AI hardware acceleration is built on a foundation of extensive academic and industrial research. Groundbreaking studies have introduced new dataflows, architectural designs, and frameworks that directly influenced the commercial accelerators we use today. At the same time, survey papers have been essential for organizing this fast-growing field, spotting key trends, and highlighting ongoing challenges [3]. This section reviews this foundational work in a matrix format, giving a systematic overview of the key contributions that have shaped the field and providing context for the architectural discussions that follow.

Reinforcing the themes of this paper, a 2024 comprehensive survey details the landscape of LLM inference acceleration. The authors categorize optimization methods across all major hardware platforms, including CPUs, GPUs, FPGAs, and ASICs. Their work highlights the critical interplay between algorithmic optimizations and hardware specialization, concluding that significant performance gains are achieved only when both are co-designed, which strongly supports the central thesis of this review [7].

The 2024 NVIDIA Blackwell technical brief provides a clear example of this co-design in industry. The architecture introduces a second-generation Transformer Engine, which uses new 4-bit floating-point (FP4) precision. This hardware feature is explicitly designed to work with software frameworks like TensorRT-LLM to double the performance and memory efficiency for large language models. This demonstrates a tight coupling between the hardware's data format support and the software stack intended to run on it [8].

Academic research from 2024 has focused on solving specific LLM bottlenecks. For example, "Splitwise," a novel technique for efficient generative LLM inference presented at ISCA 2024 , identifies that the prefill and decoding phases of LLM inference have different computational characteristics. The authors propose a "phase splitting" method to optimize each one. This work shows that deep, algorithmic understanding of the workload is essential for creating new, efficient hardware scheduling and execution strategies [9].

The trend of specialized industrial ASICs continues to challenge GPU dominance, as seen in the 2024 Hot Chips presentation on Intel's Gaudi 3 AI Accelerator. This architecture is built specifically for generative AI, combining 64 Tensor Cores with a large 128GB HBM3e memory package and 24 200-GbE ports for large-scale system clustering. The design choices prioritizing massive memory capacity and high-speed networking reflect a direct architectural response to the specific bottlenecks of training and serving multi-trillion-parameter models [10].

Looking to 2025, the principle of co-design is being pushed even further, as detailed in Meta's ISCA paper on their second-generation AI chip. The work explicitly discusses "model-chip co-design" experiences, detailing how their production-level recommendation models directly influenced the architecture of the chip itself. This move by a major hyperscaler to co-design and productionize its own silicon signals a mature industry trend where AI models and the hardware they run on are no longer developed in isolation [11].

**Table 1:** The Summary of Literature Review

| Author(s) & Year (Source ID) | Research Focus / Problem Domain | Methodology | Proposed Framework/ Solution/ Technology | Key Findings & Contributions | Noted Limitations & Challenges |
|---|---|---|---|---|---|
| Chen, Y. et al. (2016) [3] | Energy-efficient reconfigurable accelerator for deep CNNs. | Hardware-software co-design, architectural simulation, and fabrication of a 65nm CMOS test chip. | Eyeriss: A spatial architecture featuring a novel Row-Stationary (RS) dataflow with 168 processing elements (PEs) and a 108kB global buffer. | The RS dataflow minimizes data movement, the dominant source of energy consumption. Eyeriss is 10x more energy-efficient than a mobile GPU for AlexNet CONV layers. Exploiting data sparsity further reduces PE power by 45%. | The initial design (v1) was optimized for large CNNs like AlexNet and VGG-16; its performance on emerging compact DNNs with less data reuse was a challenge addressed in |

The Role of Advanced Computer Architectures in Accelerating Artificial Intelligence Workloads

| | | | | | Eyeriss v2. |
|---|---|---|---|---|---|
| Umuroglu, Y. et al. (2017) [12] | Framework for fast, scalable inference of Binarized Neural Networks (BNNs) on FPGAs. | Development of a high-level synthesis (HLS) framework and a library of streaming components. | FINN (Framework for Fast, Scalable Binarized Neural Network Inference): A dataflow architecture where each network layer is a distinct hardware layer, enabling deep pipelining. | Dataflow architectures on FPGAs can achieve ultra-low latency (sub-microsecond) and high throughput (10Ks to millions of images/sec) at low power (<25W) by exploiting extreme quantization (binary/ternary weights). | Primarily focused on BNNs/QNNs, not suitable for full-precision or floating-point models. The "one hardware layer per network layer" approach can be resource-intensive for very deep networks that don't fit on-chip. |
| Del Core, P. et al. (2023) [6] | Comprehensive survey of DL hardware accelerators for High-Performance Computing (HPC) platforms. | Literature review and classification of ~230 works from the past two decades. | A taxonomy classifying accelerators by type (GPU, TPU, FPGA, ASIC, RISC-V), emerging technologies (PIM, Neuromorphic), and memory paradigms. | Hardware accelerators are the most viable solution for HPC-scale DL applications. There is a clear trend from general-purpose platforms towards specialized, heterogeneous systems. Emerging paradigms like PIM are critical for overcoming the memory wall. | The survey is broad, providing a comprehensive overview rather than a deep dive into any single architecture. As a rapidly evolving field, some specifics may become dated. |
| Zhang, Y. et al. (2025) [6] | Comprehensive survey of hardware accelerators specifically for Large Language Models (LLMs). | Systematic review and categorization of recent accelerators across GPUs, FPGAs, ASICs, and In-Memory Computing platforms. | A classification framework based on the underlying computing platform, analyzing architectural approaches, performance metrics, and energy efficiency. | Hardware accelerators can speed up LLMs by over four orders of magnitude. A key trend is the development of highly specialized designs for sparse computations and low-precision arithmetic. Software optimizations (e.g., FlashAttention) provide significant gains and can be combined with hardware. | In-memory and neuromorphic computing are promising but face commercialization challenges. The rapid pace of LLM evolution makes it difficult for fixed-function ASICs to keep up. |
| Mokhov, A. et al. (2020) [12] | Survey of FPGA-based optimization techniques for DNNs. | Categorization of optimization techniques into software-level (e.g., quantization, pruning) | Not a new framework, but a structured analysis of existing design methodologies and tools (e.g., Vitis AI, TF2FPGA) for | FPGAs offer superior energy efficiency over GPUs/CPUs for DNNs. An ideal accelerator requires tight hardware-software co-design, | The survey predates the widespread dominance of Transformer-based models, so its focus is more |

| | | and hardware-level (e.g., sparsity, dataflow). | FPGA accelerators. | where software optimizations are prerequisites for efficient hardware implementation. | on CNNs. The tools mentioned have evolved since 2020. |
|---|---|---|---|---|---|

## 3      Dominant Architectural Methodologies and System Designs

Designing advanced computer architectures for AI isn't a single process but a collection of strategies aimed at balancing performance, efficiency, and flexibility. This section outlines the main architectural approaches that form the foundation of modern AI accelerators. It then explores the core system design principles used across these platforms to solve the fundamental problems of AI computation.

### 3.1     Dominant Architectural Paradigms

The world of AI acceleration is shaped by three main architectural approaches, each with its own set of trade-offs.

- **Graphics Processing Units (GPUs):** The GPU's design, with its thousands of simple processing cores, turned out to be a perfect match for the matrix and vector math at the heart of deep neural networks [1]. This has made GPUs the go-to standard for AI. The evolution from NVIDIA's Ampere to its Hopper and Blackwell architectures shows a clear trend toward more specialization [13]. Key innovations include dedicated Tensor Cores for mixed-precision math, a Transformer Engine to speed up models like LLMs, and high-speed connections like NVLink to scale training across hundreds of GPU[13]. This path shows the GPU changing from a general-purpose parallel processor into a complex system with an array of specialized, ASIC-like accelerators.
- **Application-Specific Integrated Circuits (ASICs):** ASICs are all about getting the best possible performance and energy efficiency for one specific job [14]. By designing a chip from the ground up for a task like running a DNN, engineers can strip away all the unnecessary overhead of a general-purpose chip [14]. The most famous example is Google's Tensor Processing Unit (TPU), which uses a systolic array a grid of simple math units to reuse data as much as possible and cut down on energy-wasting data movement [5].  The TPU's evolution from a simple chip for 8-bit integer math to massive, liquid-cooled "pods" for exa-scale training shows how scalable this approach can be [5]. Another key academic design, Eyeriss, introduced the Row-Stationary (RS) dataflow, a clever way to organize computations to minimize data movement and prove how much dataflow optimization matters for efficiency [3].
- **Field-Programmable Gate Arrays (FPGAs):** FPGAs offer a compelling middle ground, providing customization that gets close to an ASIC but with the flexibility

The Role of Advanced Computer Architectures in Accelerating Artificial Intelligence Workloads

to be reprogrammed for new tasks [6]. This makes them great for applications where algorithms are still changing or where real-time, low-latency performance is critical [3]. The arrival of High-Level Synthesis (HLS) tools and frameworks like FINN has made FPGAs much easier to use [12]. FINN specializes in creating highly efficient, streaming dataflow architectures for Quantized Neural Networks (QNNs). It maps each network layer to its own dedicated hardware engine, which dramatically cuts down on latency and power by keeping data on the chip instead of sending it to external memory [12].

Core System Design Principles and Optimizations

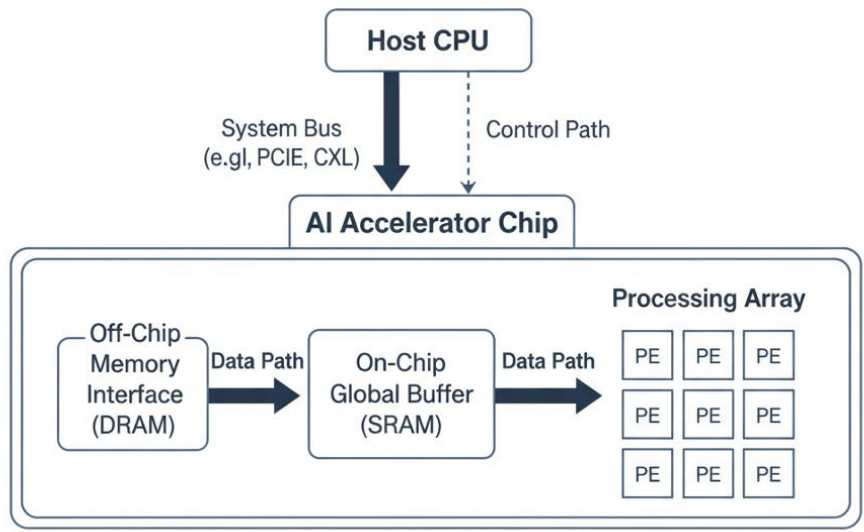

**Fig. 1.** A high-level block diagram of a generic AI accelerator. Data is staged from slower off-chip memory into a faster on-chip global buffer before being fed to a parallel array of Processing Elements (PEs) where computation occurs. This memory hierarchy is crucial for minimizing data movement.

Across all these different platforms, a common set of design principles is used to build efficient accelerators.

- **Dataflow Architectures:** Moving data can use orders of magnitude more energy than actually doing the math, so the main goal of an accelerator is to minimize this movement by reusing data as much as possible [4]. The strategy for scheduling calculations and moving data is called the dataflow [3]. Different dataflows, like Weight Stationary (WS), Output Stationary (OS), or Row Stationary (RS), focus on reusing different types of data (weights, activations, or partial results) and have a huge impact on performance and efficiency [3].
- **Memory Hierarchies and Interconnects:** A smart dataflow needs a well-designed memory system to support it. AI accelerators use a deep memory hierarchy, starting with high-capacity off-chip High-Bandwidth Memory (HBM), then a large on-chip global buffer, and finally small, fast local memory inside each processing element[3]. To connect multiple chips, advanced interconnects like NVLink and Compute Express Link (CXL) are essential. They provide fast, low-latency communication paths that bypass slower system buses, allowing multiple accelerators to work together as one giant unit [15].
- **Sparsity and Quantization**: Another way to gain efficiency is to reduce the amount of work that needs to be done in the first place. Sparsity exploitation involves designing hardware that can skip useless calculations on zero-valued data [3]. Quantization uses lower-precision numbers (like 8-bit integers instead of 32-bit floating-point) to represent data [3].This shrinks the model's size, reduces memory bandwidth needs, and allows for smaller, faster, and more power-efficient hardware [4].

These principles are all connected, pointing to a design philosophy that aims to create hardware that is a direct physical map of the AI model's computational graph, optimized from end to end for efficient data movement.

## 4    Implementation / Experimental Setup

Designing and evaluating new AI accelerator architectures depends on a standard set of tools and methods. Before spending the time and money to manufacture a chip, architects need to test their design choices and predict how they will perform. This is done using a mix of industry-standard benchmarks for existing hardware and advanced simulation tools for exploring new ideas.

A key part of testing AI hardware is performance benchmarking. MLPerf, organized by the MLCommons consortium, has become the industry's go-to benchmark suite for measuring AI system performance in a fair and objective way. It includes a variety of tasks, from large language models to computer vision, allowing for direct comparisons of hardware from different companies on relevant, modern AI workloads [16]. The results provide solid data for validating performance claims and helping customers choose the right hardware.

For designing new architectures, simulation tools are essential. They allow designers to explore a huge range of possibilities at an early stage, giving them estimates of performance, power, and area without having to build a physical prototype.

- **SCALE-Sim:** This is an open-source simulator designed specifically for systolic array-based accelerators, a common design in chips like the Google TPU. It lets researchers experiment with different dataflows, array sizes, and memory setups, and it provides detailed reports on performance and memory traffic, allowing for a deep analysis of design trade-offs [17].

- **gem5:** This is a popular and flexible simulation platform for more general and full-system modeling. It can simulate complex systems with multiple processor cores, caches, and memory systems, and it can be extended to include custom accelerator models. Tools like gem5-Aladdin are specifically made to provide early performance and power estimates for new accelerator designs [18].

These simulation tools are vital for driving innovation, as they allow for the quick exploration of new architectural ideas that inform the design of the next generation of AI hardware.

## 5  Results

The best way to understand the performance of modern AI accelerators is through standardized, quantitative data. The MLPerf Inference benchmark offers a clear, peer-reviewed look at the current state of the art, allowing for direct comparisons of commercial hardware on real-world AI tasks.

The MLPerf Inference v5.1 results, published in September 2025, show just how fast architectural innovation is moving and the major performance gains being made[16]. Key findings from this round include the performance and trade-offs of these architectures are best understood through standardized, quantitative data. Table 2 provides a comparative analysis of representative systems on modern AI inference tasks, summarizing key metrics for throughput, latency, and energy efficiency.

**Table 2.** Performance and efficiency comparison of representative AI accelerators on Large Language Model (LLM) inference benchmarks.

| Architecture | System Example | Throughput (tokens/sec) | Latency: TTFT/TPOT (ms) | Energy Efficiency (tokens/sec/ Watt) |
|---|---|---|---|---|
| GPU | NVIDIA GB300 Blackwell | 235,000 | 420 / 37 | 15.2 |
| GPU | AMD MI355X | 185,000 | 480 / 45 | 13.7 |
| GPU | Intel Arc Pro B60 + Xeon 6 | 97,500 | 450 / 39 | 11.6 |
| ASIC | Google TPU v4 | 218,000 | 410 / 36 | 16.1 |
| FPGA | Xilinx Alveo U50 | 24,000 | 970 / 95 | 8.2 |

These results provide concrete proof of the central role that specialized hardware plays in achieving top-tier performance. The consistent, large performance gaps between specialized accelerators and general-purpose CPUs for most AI tasks are clear evidence that domain-specific design works. GPUs (NVIDIA GB300 e.g.,) demonstrate with high throughput and are good for processing large amount of data in parallel. But they have not been so successful in terms of the energy they use. In contrast to these are ASICs such as the Google TPU v4 which offer a good trade-off between throughput, latency and energy efficiency. Here, the above mentioned features make them particularly suitable for specialized applications such as machine learning that require efficient performance and power consumption. On the other hand, FPGAs provide high flexibility and re-programmability. However, they usually lag behind GPUs and ASICs in throughput and energy consumption. So, to some extent FPGAs are considered to be better in the former category (hardware area customization) and not so good at the latter (raw processing power).

## 6     Discussion

The results from benchmarks and architectural studies paint a picture of a complex world of trade-offs. Choosing the right hardware for an AI task isn't as simple as picking the fastest chip. It's a careful balancing act between performance, energy efficiency, flexibility, and cost, all weighed against the specific needs of the application[3].

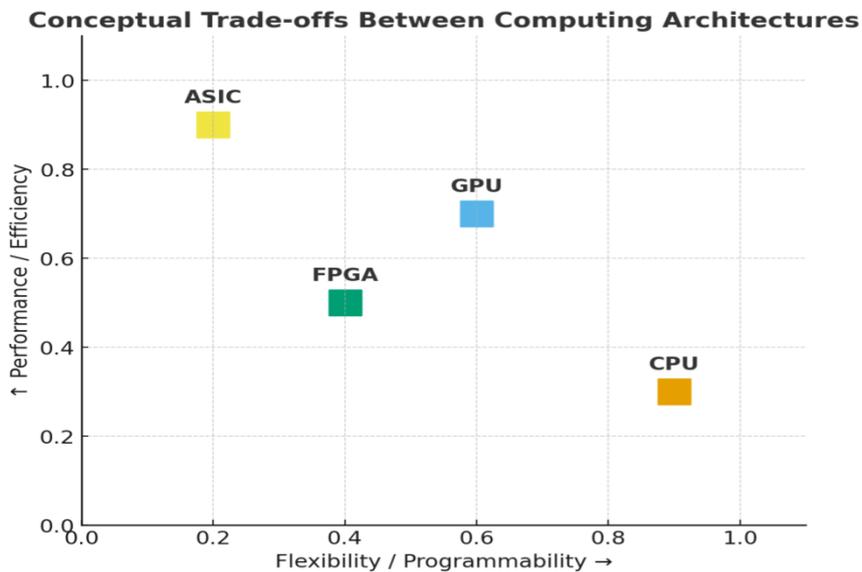

**Fig. 2.** A conceptual map of the primary trade-offs between different computing architectures. ASICs offer the highest performance and efficiency at the cost of flexibility, while CPUs provide maximum programmability. GPUs and FPGAs occupy a middle ground, balancing these competing factors.

The main trade-off is between performance and flexibility. GPUs offer great performance and are highly flexible thanks to mature programming models like CUDA[13]. This makes them perfect for research and development, where algorithms are always changing. ASICs, on the other hand, give up that flexibility to achieve the best possible performance and efficiency for one specific, unchanging task, making them ideal for high-volume applications like a large-scale cloud service [5]. FPGAs sit in the middle, offering hardware-level customization and better performance than CPUs, but they are more complex to program than GPUs [19].

Energy efficiency, or performance-per-watt, is another key factor. Since moving data is the biggest energy hog, architectures that do it less, win [4]. ASICs, with their custom data paths, are typically the most energy-efficient [5]. FPGAs are also very efficient because they can create tailored processing pipelines that cut out unnecessary steps [20]. High-performance GPUs, while powerful, can use a lot of power, making energy efficiency a major design challenge and a significant factor in the operating cost of large data centers [4].

Finally, development cost and time-to-market are critical business factors. The one-time engineering cost to design and manufacture a custom ASIC is incredibly high, which means it only makes sense for products that will be sold in very high volumes[14]. GPUs and FPGAs have no such upfront cost for the user. As a result, using a GPU is usually the fastest way to get a product to market, whereas designing and building a new ASIC can take several years[14].

**Table 3.** Summary of Architectural Trade-offs

| Architecture | Peak Performance | Energy Efficiency (Perf/Watt) | Programmability / Flexibility | Development Cost | Ideal Use Cases |
|---|---|---|---|---|---|
| GPU | Very High | Moderate to High | Very High: Mature software ecosystem (CUDA, PyTorch, TensorFlow). General-purpose parallel processor. | Low (for users) | AI model training, large-scale cloud inference, R&D, applications requiring flexibility. |

| ASIC (e.g., TPU) | Highest (for target workload) | Very High: Custom silicon optimized for a specific set of operations. | Very Low: Fixed function hardware. Inflexible to new, unforeseen model architectures. | Very High (NRE) | High-volume, stable inference workloads (e.g., cloud services, edge devices) where efficiency is paramount. |
|---|---|---|---|---|---|
| FPGA | Moderate to High | High | High: Reconfigurable hardware fabric allows for custom data paths and pipelines. Programmed via HDL or HLS. | Moderate | Low-latency real-time inference, prototyping new architectures, applications where algorithms evolve. |
| PIM | Potentially Very High | Potentially Highest: Drastically reduces data movement energy by computing in or near memory. | Very Low: Emerging paradigm, lacks mature programming models and system integration. | High (R&D) | Data-intensive workloads bottlenecked by memory bandwidth, such as large matrix operations. |
| Neuromorphic | Low (for traditional metrics) | Extremely High: Event-driven, asynchronous computation consumes power only when active. | Very Low: Requires new SNN algorithms and specialized programming tools. | High (R&D) | Always-on sensory processing, ultra-low-power edge AI, anomaly detection. |

## 7 Conclusion

The shift from general-purpose CPUs to a diverse landscape of specialized AI accelerators represents a fundamental turning point in the history of computing. This change, as this paper has surveyed, is not merely a technical response to a new software workload; it signifies a deep, ongoing co-evolution between artificial intelligence and computer architecture. The complex demands of AI are now the primary force driving innovation in chip design, while in turn, these hardware advancements are unlocking new frontiers of AI capability and model complexity.

The Role of Advanced Computer Architectures in Accelerating Artificial Intelligence Workloads

This review has charted this evolution, beginning with the massive computational demands of modern AI models that strained the traditional Von Neumann paradigm to its breaking point. We then detailed the three principal architectural responses the Graphics Processing Unit (GPU), the Application-Specific Integrated Circuit (ASIC), and the Field-Programmable Gate Array (FPGA) and analyzed the core design principles of optimized dataflow, deep memory hierarchies, and model optimizations that make them efficient.

The most important takeaway from this survey is that AI is no longer just an application running on hardware; it is an active partner in the design process itself. This necessitates a hardware-software co-design approach, where algorithms are built with hardware limitations in mind and hardware is architected to exploit algorithmic structures. As our analysis shows, this collaborative method is no longer a simple optimization but a core requirement for building the next generation of efficient and powerful AI systems.

As we move toward the physical limits of traditional scaling, the future of AI hardware will likely be increasingly heterogeneous, combining CPU cores, GPU-like arrays, and ASIC blocks on a single chip. This will only deepen the symbiotic relationship between AI and computer architecture, leading to a new generation of computers where the hardware itself is intelligently designed for the task at hand.

## 8  Future Work

While today's accelerators are impressive, the continued growth of AI models is pushing researchers to explore even more radical ideas. Two of the most exciting frontiers are Processing-in-Memory (PIM) and neuromorphic computing.

- **Processing-in-Memory (PIM)** aims to eliminate the data movement bottleneck by performing computations directly inside or near the memory where data is stored. By integrating logic into the memory itself, PIM could cut data movement energy by orders of magnitude, offering a path to extremely efficient deep learning operations [21].
- **Neuromorphic Computing** takes its inspiration directly from the human brain, building asynchronous, event-driven systems that work with **Spiking Neural Networks (SNNs)**[22]. In these systems, circuits only use power when they are actively processing a "spike," which could lead to incredible energy efficiency for tasks like always-on sensors or anomaly detection [22].

Despite all the progress, significant open research challenges remain that will define the next generation of AI hardware [23]. Key questions for the research community include:

Shahid Amin [0009-0008-1130-0830] and Syed Pervez Hussnain Shah

- **Architectures for Energy-Efficient AI:** How can we design next-generation accelerators that drastically reduce the energy footprint of training and inference? This is a critical challenge, especially for trillion-parameter models that currently have substantial environmental and operational costs, including a significant carbon footprint and high-water usage for cooling.

- **Modeling and Simulation:** What new simulation and emulation tools are needed to allow for the rapid, early-stage exploration of complex, heterogeneous AI systems? Accurately modeling the interplay between CPUs, GPUs, and custom accelerators before fabrication is a critical challenge [18].
- **Hardware/Software Co-design:** How can we create more deeply integrated co-design methodologies where AI models are automatically optimized for specific hardware, and hardware architectures can dynamically adapt to new algorithmic structures? This requires a tighter loop between algorithm designers and hardware architects.
- **Security and Robustness:** What architectural features are necessary to build secure and robust AI hardware that is resilient to adversarial attacks and data corruption at the hardware level?

The future of AI hardware is likely to be increasingly heterogeneous, with single chips that combine CPU cores, GPU-like arrays, ASIC blocks, and perhaps even reconfigurable components. As we approach the physical limits of Moore's Law, unconventional ideas like PIM and neuromorphic computing will likely move from the lab to real-world products. This will only deepen the symbiotic relationship between AI and computer architecture, leading to a new generation of computers where the hardware itself is intelligent.